\documentclass[aps,preprint]{revtex4}%
\usepackage{amsfonts}
\usepackage{amsmath}
\usepackage{amssymb}
\usepackage{graphicx}%
\providecommand{\U}[1]{\protect\rule{.1in}{.1in}}

\begin{document}
\preprint{ }
\title[critical current enhancement in YBCO]{Enhancement of the Critical Current Density of YBa$_{2}$Cu$_{3}$O$_{x}$
Superconductors under Hydrostatic Pressure}
\author{T. Tomita and J.S. Schilling}
\affiliation{Department of Physics, Washington University, CB 1105, One Brookings Dr, St.
Louis, MO 63130}
\author{L. Chen, B.W. Veal, and H. Claus}
\affiliation{Materials Science Division, Argonne National Laboratory, 9700 S. Cass Avenue,
Argonne, IL 60439}
\keywords{}
\pacs{74.62.Fj, 74.25.Sv, 74.72.Bk, 74.81.Bd}

\begin{abstract}
The dependence of the critical current density $J_{c}$ on hydrostatic pressure
to 0.6 GPa is determined for a single 25$^{\circ}$ [001]-tilt grain boundary
in a bicrystalline ring of nearly optimally doped melt-textured YBa$_{2}%
$Cu$_{3}$O$_{x}$. $J_{c}$ is found to increase rapidly under pressure at +20
\%/GPa. A new diagnostic method is introduced (pressure-induced $J_{c}$
relaxation) which reveals a sizeable concentration of vacant oxygen sites in
the grain boundary region. Completely filling such sites with oxygen anions
should lead to significant enhancements in $J_{c}.$

\end{abstract}
\date{November 17, 2005}
\maketitle

One of the primary factors limiting applications using ceramic high-$T_{c}$
superconductors has been the inability of grain boundaries (GBs) to tolerate
large critical current densities ($J_{c}\simeq10^{6}-10^{7}$ A/cm$^{2}),$
particularly at high magnetic field levels \cite{hilgenkamp1}. The value of
$J_{c}$ in polycrystals can be dramatically enhanced by reducing GB mismatch
angles to values less than 4$^{\circ}$ through suitable grain alignment
(texturing) procedures \cite{dimos1}\ and by properly preparing the GB region
\cite{hilgenkamp1}, for example, through Ca doping \cite{hammerl1}. Very
recent detailed TEM and EELS studies have attributed the success of Ca doping
both to strong Ca segregation near the dislocation cores \cite{song1} and to
reduction in strains in the GB region which reduces the depletion of oxygen
anions, thus enhancing the (hole) carrier concentration in the GB
\cite{klie1}. In either scenario, oxygen depletion in the GB leads to serious
reductions in $J_{c}.$

The oxygen concentration in the GB region is evidently a parameter of vital
importance in the optimization of $J_{c},$ even as the oxygen content within a
grain plays a major role in the optimization of $T_{c}$. In melt-textured
YBa$_{2}$Cu$_{3}$O$_{x}$ (YBCO) bicrystals we have observed that the GB
$J_{c}$ increases monotonically as the oxygen content in the bulk is increased
from the underdoped condition to nearly full oxygenation at $x=6.98 $
\cite{veal2}, even though $T_{c}$ passes through a maximum at optimal doping
$(x\approx6.95).$ This behavior was also observed in thin film GBs
\cite{claus3}. The determination of the concentration of vacant oxygen sites
in the GB is, however, a difficult problem \cite{klie1}. Even when the bulk
material is fully oxygenated, the GBs are likely oxygen deficient.

In this paper we introduce an experimental procedure, pressure-induced $J_{c}
$ relaxation, which provides a sensitive test for the presence of oxygen
vacancies in the GB region. In the slightly overdoped YBCO bicrystal studied,
the single 25$^{\circ}$ GB is found to be clearly oxygen deficient.

Early experiments on YBa$_{2}$Cu$_{4}$O$_{8}$ and Tl$_{2}$CaBa$_{2}$Cu$_{2}%
$O$_{8+\delta}$ ceramics \cite{diederichs1} and polycrystalline YBCO thin
films \cite{budko1} indicated that the bulk $J_{c}$ increases with pressure,
but relaxation behavior was not studied. Samples containing a single
well-defined grain boundary are clearly needed for quantitative tests of GB
models. Such experiments have recently become possible through the
availability of YBCO bicrystalline melt-textured rings with single [001]-tilt
GBs with varying misorientation angle $\theta$ and oxygen content $x$
\cite{claus1}.

For a single 25$^{\circ}$ GB in a YBCO bicrystal, we find that $J_{c}$
increases rapidly with hydrostatic pressure. The rate of increase ($+20$
\%/GPa) is much more rapid than that expected from a simple GB model which
takes into account the compression of the tunnel barrier width $W$ alone.

The Vespel sample holder containing the YBCO bicrystalline ring (4 mm OD
$\times$ 2 mm ID $\times$ 0.5 mm) is placed in the 7 mm dia bore of a He-gas
pressure cell (Unipress). Two counterwound Cu pickup coils are positioned on
the outside of the sample holder for the $ac$ susceptibility measurement. To
avoid heating effects at high field amplitudes ($\sim$ 300 G), the field coil
is placed \textit{outside} the tailpiece of the cryostat and the $ac$
frequency is reduced to 1 Hz to ensure full field penetration to the sample.
Standard $ac$ susceptibility techniques are used with a SR830 digital lock-in
amplifier. Further details of the high pressure \cite{tomita1,tomita2} and
sample preparation procedures for the melt-textured YBCO bicrystalline rings
\cite{claus1} are given elsewhere.

In previous studies the GB $J_{c}$ was determined from the change in
magnetization of YBCO bicrystalline rings as a function of temperature and
$dc$ magnetic field using a SQUID magnetometer \cite{claus1}. Here we utilize
$ac$ susceptibility measurements to obtain similar information. In Fig 1
(left) an $ac$ field $H(t)=H_{ac}\sin\omega t$ with amplitude $H_{ac}=$ 10 G
at 1 Hz is applied by the primary coil to the ring sample inducing ring
currents $I(t)=I_{ac}\cos\omega t$. In the single-turn solenoid approximation,
we have $I_{ac}=DH_{ac},$ $D$ being the average of the inner and outer ring
diameters \cite{claus1}. As seen in Fig 1 (left), for temperatures below 67 K
at ambient pressure, the oscillatory magnetic field generated by these ring
currents is equal and opposite to the oscillatory applied field, thus
preventing flux from entering. As the temperature is increased above 67 K,
however, the ring current $I_{ac}$ exceeds the critical current $I_{c}$ and
magnetic flux begins to\ flow in and out of the ring through the weaker of the
two grain boundaries. Knowing the cross-sectional area $A$ of the ring, one
can estimate the critical current density $J_{c}=I_{c}/A.$ For temperatures in
the region 85 - 91 K, $\chi^{\prime}(T)$ reaches a plateau where the applied
flux is only excluded from the superconducting material in the ring itself.
Finally, above $T_{c}\simeq$ 92 K the ring is in the normal state and magnetic
flux penetrates uniformly through the entire ring.

From this measurement of $\chi^{\prime}$ versus temperature, therefore, two
important properties of the nearly optimally doped YBCO ring are determined:
\ (1) the value of the critical current density $J_{c}(T_{kink})\simeq
DH_{ac}(T_{kink})/A$ through the GB at the temperature of the kink in
$\chi^{\prime}$ and (2) the value of the transition temperature $T_{c}%
\simeq91.84$ K\emph{\ }(midpoint of $\chi^{\prime}$) in bulk YBCO. If the
oscillatory field amplitude $H_{ac}$ is increased, $T_{kink}$ shifts to lower
temperatures, yielding the temperature dependence of $H_{ac}(T)$ or $J_{c}(T)$
shown in Fig 1 (right). The $J_{c}(T)$ dependences obtained in these $ac$
susceptibility studies are in good agreement with the results of magnetization
measurements in $dc$ fields using a SQUID magnetometer\emph{\ }\cite{veal2},
as pointed out earlier by Herzog \textit{et al} \cite{herzog1}.

Fig 1 (left) also displays $\chi^{\prime}(T)$ data obtained after the
application of 0.6 GPa hydrostatic He-gas pressure at ambient temperature.
$T_{kink}$ is seen to shift to higher temperatures, indicating that $J_{c}$
increases under pressure. This result is brought out clearly by the $J_{c}(T)
$ data in Fig 1 (right) at both ambient and 0.6 GPa pressure; these data are
fit using the expression%
\begin{equation}
J_{c}(T)=J_{c}(0)\left[  1-T/T_{c}\right]  ^{\beta},\label{1}%
\end{equation}
where $\beta=0.89$ and $J_{c}(0)=1.44$ and 1.61 kA/cm$^{2}$ for $P=0$ and 0.6
GPa, respectively. For temperatures below $T_{c},$ the relative pressure
dependence $J_{c}^{-1}(dJ_{c}/dP)=d\ln J_{c}/dP\simeq+0.20$ GPa$^{-1}(+20$
\%/GPa) is obtained$.$ This value increases to $+0.26$ GPa$^{-1}$ if a $dc$
magnetic field of 120 G is applied above 92 K before cooling.\emph{\ }Note
that the superconducting transition temperature decreases slightly under
pressure ($d\ln T_{c}/dP\simeq-25\times10^{-4}$ GPa$^{-1}$), a rate $\sim$
$80\times$ less than that of $J_{c}.$ Strong pressure-induced $J_{c}$
enhancements are also found for nearly optimally doped and underdoped YBCO
bicrystals with mismatch angles 4$^{\circ}$ to 31$^{\circ}$, as discussed
elsewhere \cite{tomita1,tomita2}.

The rapid increase in $J_{c}$ under pressure suggests that $J_{c}$ can be
further increased in applications at ambient pressure by compressing textured
YBCO material through suitable processing procedures. The relative change in
$J_{c}$ with GB width $W$ is given by $d\ln J_{c}/d\ln W=$ $\kappa_{GB}%
^{-1}(d\ln J_{c}/dP),$ where the compressibility of the GB is defined by
$\kappa_{GB}\equiv-d\ln W/dP.$ If we assume to a first approximation that
$\kappa_{GB}$\ is roughly comparable with the average linear compressibility
$\kappa_{a}\equiv-d\ln a/dP$ of YBCO in the CuO$_{2}$ plane \cite{note1},
where $\kappa_{a}\approx2\times10^{-3}$ GPa$^{-1}$ \cite{jorgensen1} and $a$
is an in-plane lattice parameter, it follows that $d\ln J_{c}/d\ln
W\approx-100.$ This implies that $J_{c}$ increases under pressure at a rate
which is 100$\times$ more rapid than the decrease in GB width! Compressing the
GB by 10\% should, therefore, lead to a tenfold enhancement in $J_{c}.$
Lattice compression by a few \% can be readily achieved through epitaxial
growth techniques.

We now examine the question of the mechanism behind the strong enhancement in
$J_{c}$ as the GB is compressed. High pressures may modify the GB in a number
of different ways, including: \ (1) reduction of the tunneling barrier width
$W$ and change in the tunneling barrier height $\phi$, (2) promotion of oxygen
ordering in the GB, in analogy with the well studied pressure-induced oxygen
ordering effects in the bulk \cite{klehe1,sadewasser1,fietz1,chu1}.

Considering the first possibility, the WKB approximation applied to a
potential barrier gives the following simple expression: $\ J_{c}=J_{co}%
\exp(-2KW),$\emph{\ }where $W$ is the barrier width, $J_{co}$ is the critical
current density for zero barrier width, i.e. no grain boundary, and
$K=\sqrt{2m\phi}/\hbar$ is the decay constant which increases with the barrier
height $\phi$ \cite{browning1}. Since parallel studies on melt-textured YBCO
rings with no GB reveal that $d\ln J_{co}/dP\simeq0$ \cite{tomita1}, one
obtains for the relative pressure dependence of $J_{c}$%
\begin{equation}
\frac{d\ln J_{c}}{dP}\simeq-\left[  \left(  \frac{d\ln W}{dP}\right)
\ln\left(  \frac{J_{co}}{J_{c}}\right)  \right]  -%
\frac12
\left[  \left(  \frac{d\ln\phi}{dP}\right)  \ln\left(  \frac{J_{co}}{J_{c}%
}\right)  \right]  .\label{2}%
\end{equation}
Does the strong enhancement in $J_{c}$ under pressure arise primarily from the
first term on the right in Eq. (2), i.e. from the decrease in the barrier
width $W$? To address this question,\emph{\ }we assume \cite{note1}, as above,
that to a first approximation -$d\ln W/dP\equiv\kappa_{GB}\approx\kappa
_{a}\approx2\times10^{-3}$\emph{\ }GPa$^{-1}$\emph{.} From Fig 1 we see that
for the $\theta=25^{\circ}$ ring $J_{c}\simeq1,400$ A/cm$^{2}$ at low
temperatures and ambient pressure. Setting $J_{co}\approx250,000$ A/cm$^{2}$
for a melt-textured YBCO ring with no GB \cite{tomita1}, we find ($-d\ln
W/dP)\ln(J_{co}/J_{c})\approx+0.01$ GPa$^{-1},$ a value $20\times$ less than
the above experimental value $d\ln J_{c}/dP\simeq+0.20$ GPa$^{-1}$. Within the
simple tunneling barrier model, therefore, the strong enhancement in $J_{c}%
$\ under pressure does\ not arise from the compression of the GB width
$W,$\ but may\ be the result of a strong reduction in the barrier height
$\phi$\ at the GB. From Eq. (2) a pressure derivative of only\emph{\ \ }%
$d\ln\phi/dP\simeq-0.073$ GPa$^{-1}$ would be sufficient to yield the
experimental value\ $d\ln J_{c}/dP\simeq+0.20$ GPa$^{-1}$.

The second potential mechanism proposed above for the rapid increase of
$J_{c}$ with pressure is through oxygen ordering effects in the GB. It is well
established that pressure-induced oxygen ordering effects in high-$T_{c} $
oxides can have a dominant influence on the pressure dependence of bulk
properties, such as the value of $T_{c}$ \cite{klehe1,sadewasser1,fietz1,chu1}%
. In an overdoped Tl$_{2}$Ba$_{2}$CuO$_{6+x}$ sample, for example, the
hydrostatic pressure dependence even changes sign from $dT_{c}/dP\simeq-8.9$
K/GPa to $+0.35$ K/GPa depending on whether, respectively, oxygen-ordering
effects occur or not \cite{klehe1}. Significant oxygen ordering effects in
$dT_{c}/dP$ have also been observed in YBCO \cite{sadewasser1,fietz1}.

Oxygen ordering effects in many cuprate oxides have their origin in the
appreciable mobility of oxygen anions in certain regions of the oxygen
sublattice, even at ambient temperatures. In YBCO at ambient temperatures and
below, oxygen anions in the CuO chains possess considerable mobility, in
contrast to oxygen anions in the CuO$_{2}$ planes. The degree of local order
assumed by the mobile oxygen anions changes as a function of pressure or
temperature: \ raising the temperature above ambient reduces the order in an
equilibrated system, applying pressure at ambient temperatures enhances the
order. The existence of pressure-induced oxygen ordering effects can be
readily demonstrated by applying pressure at ambient temperature to enhance
local order, but releasing it at temperatures sufficiently low (%
$<$
200 K for YBCO) to prevent the oxygen anions from diffusing back, thus
effectively freezing in the higher degree of order.

The degree of local oxygen ordering influences indirectly the charge carrier
concentration $n$ in the CuO$_{2}$ planes by changing the average valence of
the ambivalent cations (in YBCO these are the Cu cations in the CuO chains
\cite{veal1}, in Tl$_{2}$Ba$_{2}$CuO$_{6+x}$ the Tl cations in the Tl$_{2}%
$O$_{2}$ double layer \cite{klehe1}). Since $T_{c}$ is a sensitive function of
$n,$ the degree of oxygen ordering may have a sizeable effect on $T_{c}$, as
pointed out above for Tl$_{2}$Ba$_{2}$CuO$_{6+x}$.

There are two special cases where no pressure-induced oxygen ordering effects
are expected: \ (1) no mobile oxygen anions are present in the oxygen
sublattice, (2) the oxygen sublattice is completely filled up with oxygen
anions, so that no change in the degree of local ordering is possible. If
oxygen ordering effects are observed, therefore, this very fact implies that
the oxygen sublattice is only partially filled, i.e. empty oxygen sites are
available. The presence or absence of relaxation effects in $T_{c}$ can thus
be used as a diagnostic tool to test whether or not the lattice is able to
accommodate further oxygen anions.

We now apply this same procedure to test for oxygen ordering effects in the GB
region of a YBCO bicrystal. In Fig 2 (upper) one sees that the application of
0.6 GPa pressure at 290 K prompts $J_{c}$ at 9 K to increase from point 1 to
point 2. If the pressure is then released at temperatures below 50 K, $J_{c}$
is seen to \textit{not} decrease completely back to its initial value at point
1, but rather to remain at a higher value (point 3).\emph{\ }Remeasuring
$J_{c}$ after annealing the sample for
${\frac12}$
$h$ at successively higher temperatures results in no change in $J_{c}$
(points 3 through 9) until the annealing temperature $T_{a}$ reaches values
above $250$ K; at point 11 $J_{c}$ is seen to have fully relaxed back to its
initial value (point 1).

In a subsequent experiment (see Fig 2 (lower)) the relaxation time $\tau$ for
the GB relaxation process is estimated by annealing for different lengths of
time at a \textit{fixed} temperature (270 K) following pressure release at low
temperatures. The value of $J_{c}$ at 9 K is seen to show an exponential
time-dependent relaxation behavior which obeys the equation%
\begin{equation}
J_{c}(t)=J_{c}(\infty)-\left[  J_{c}(\infty)-J_{c}(0)\right]  \exp\left\{
-\left(  t/\tau\right)  ^{\alpha}\right\}  ,\label{3}%
\end{equation}
where $J_{c}(0)$ and $J_{c}(\infty)$ are the initial and fully relaxed values
of the critical current density, respectively. From the best fit to the data,
the estimated GB relaxation time is $\tau=5.9\ $h for $\alpha\simeq0.34.$
Using the Arrhenius law $\tau=\tau_{o}\exp[E_{a}/k_{B}T],$ the activation
energy is estimated to be $E_{a}=0.87$ eV, where we set $\tau_{o}%
=1.4\times10^{-12}s$ from previous studies \cite{veal1}. That this relaxation
in $J_{c}$\ indeed results from the motion of oxygen anions in the GB region
is supported by two experimental findings: \ (1) the relaxation in $J_{c}%
$\ takes place in the same temperature range (250 - 290 K) where oxygen
ordering phenomena in bulk YBCO are known to occur
\cite{sadewasser1,fietz1,chu1}, and (2), in agreement with the relaxation of
$T_{c}$\ in the bulk \cite{sadewasser1,fietz1,chu1}, the magnitude of the
relaxation effects in $J_{c}$\ increases substantially in underdoped
(underoxygenated) YBCO bicrystals \cite{tomita1} where the concentration of
empty oxygen sites is much higher, thus opening up many more relaxation channels.

We now examine the relaxation data in Fig 2 in detail and address
the\emph{\ }question whether or not pressure-induced oxygen ordering effects
in the GB region are primarily responsible for the large enhancement of
$J_{c}$ under pressure. In Fig 2 (upper) the initial application of 0.6 GPa
pressure at 290 K is seen to increase $J_{c}$ by 0.156 kA/cm$^{2}$ from 1.291
kA/cm$^{2}$ (point 1) to 1.447 kA/cm$^{2}$ (point 2), a 12\% increase, thus
implying $d\ln J_{c}/dP=+0.12\div(0.6$ GPa) $=+0.20$ GPa$^{-1},$ as cited
above. This pressure-induced increase in $J_{c}$ can be divided up into a
relaxation and a non-relaxation (intrinsic) contribution, i.e. $d\ln
J_{c}/dP=(d\ln J_{c}/dP)_{relax}+(d\ln J_{c}/dP)_{intr}.$ The relative
importance of these two contributions can be determined by releasing the
pressure at a temperature sufficiently low (%
$<$
250 K) to freeze out the relaxation contribution. If the relaxation
contribution strongly dominates, then the release of pressure at low
temperatures should cause little or no change in $J_{c}.$ The fact that
$J_{c}$ decreases by 0.125 kA/cm$^{2}$ upon release of pressure at
temperatures below 50 K implies that 0.125$\div$0.156 or 80\% of the large
initial increase of $J_{c}$ with pressure is due to non-relaxation effects. We
thus find that $(d\ln J_{c}/dP)_{relax}\simeq(+0.031\div1.291)\div(0.6$ GPa)
$=+0.04$ GPa$^{-1}$ and $(d\ln J_{c}/dP)_{intr}\simeq(+0.125\div
1.291)\div(0.6$ GPa) $=+0.16$ GPa$^{-1}.$ Oxygen ordering effects thus
contribute 20\% to the large increase in $J_{c}$ under pressure;
non-relaxation (intrinsic) phenomena, such as the reduction in the barrier
height $\phi,$ contribute 80\%.

Although the principal mechanism for the pressure-induced enhancement in
$J_{c}$ does not originate from oxygen ordering effects, the fact that this
relaxation component is sizeable implies that there must be a significant
number of vacant oxygen sites available in the GB region. As will be discussed
elsewhere \cite{tomita1}, a similar result is obtained for other nearly
optimally doped and underdoped YBCO rings with varying mismatch angles.

In summary, the critical current density $J_{c}$ across a 25$^{\circ}$
[001]-tilt GB in a slightly overdoped YBCO bicrystal is found to increase
strongly with hydrostatic pressure at the rate $+20$ \%/GPa. This suggests a
new procedure to enhance $J_{c}$ at ambient pressure: \ compress the GB as
much as possible through epitaxial growth or other chemical means. The rate of
increase of $J_{c}$ is far too large to be caused by the decrease in width $W$
of a tunnel barrier alone; a decrease in the barrier height $\phi$\ may be the
dominant effect.

Sizeable pressure-induced oxygen ordering effects are found to occur in the
GB, revealing that the value of $J_{c}$ is a complicated function of the
pressure/temperature history of the sample, i.e. $J_{c}=J_{c}(T,P,t),$ in
analogy with the well studied relaxation processes in the bulk where
$T_{c}=T_{c}(T,P,t)$ \cite{klehe1,sadewasser1,chu1}. These $J_{c}$ relaxation
effects are responsible for only 20\% of the total increase in $J_{c}$ with
pressure. However, the fact that these relaxation effects in $J_{c}$ occur at
all indicates that a significant concentration of\emph{\ }oxygen vacancies
must be present in the GB. Filling these vacancies by annealing the sample
under high oxygen pressure at high temperature or through electro-chemical
oxidation should lead to further enhancements in $J_{c}.$

\vspace{1cm}\noindent\textbf{Acknowledgments. \ }The research at Washington
University is supported by NSF Grant DMR-0404505 and that at the Argonne
National Labs by the U.S. Department of Energy, Basic Energy
Sciences-Materials Sciences, under Contract W-31-109-ENG-38.

\begin{center}
{\Huge Figure Captions}
\end{center}

\bigskip\ 

\noindent\textbf{Fig. 1. \ }(left) Real part of $ac$ susceptibility
$\chi^{\prime}$ versus temperature at $H_{ac}=$ 10 G (1 Hz) for nearly
optimally doped YBCO bicrystalline ring with 25$^{\circ}$ GB at 0 GPa
($\bullet$) and 0.6 GPa ($\circ$) pressure. Arrows mark temperature of kink
$T_{kink}$ in $\chi^{\prime}(T)$ where flux begins to enter ring (see
illustration at top). (right) $ac$ field amplitude $H_{ac}$ and calculated
critical current density $J_{c}(T)$ versus temperature at 0 and 0.6 GPa. Solid
lines are fits using Eq (1).\bigskip

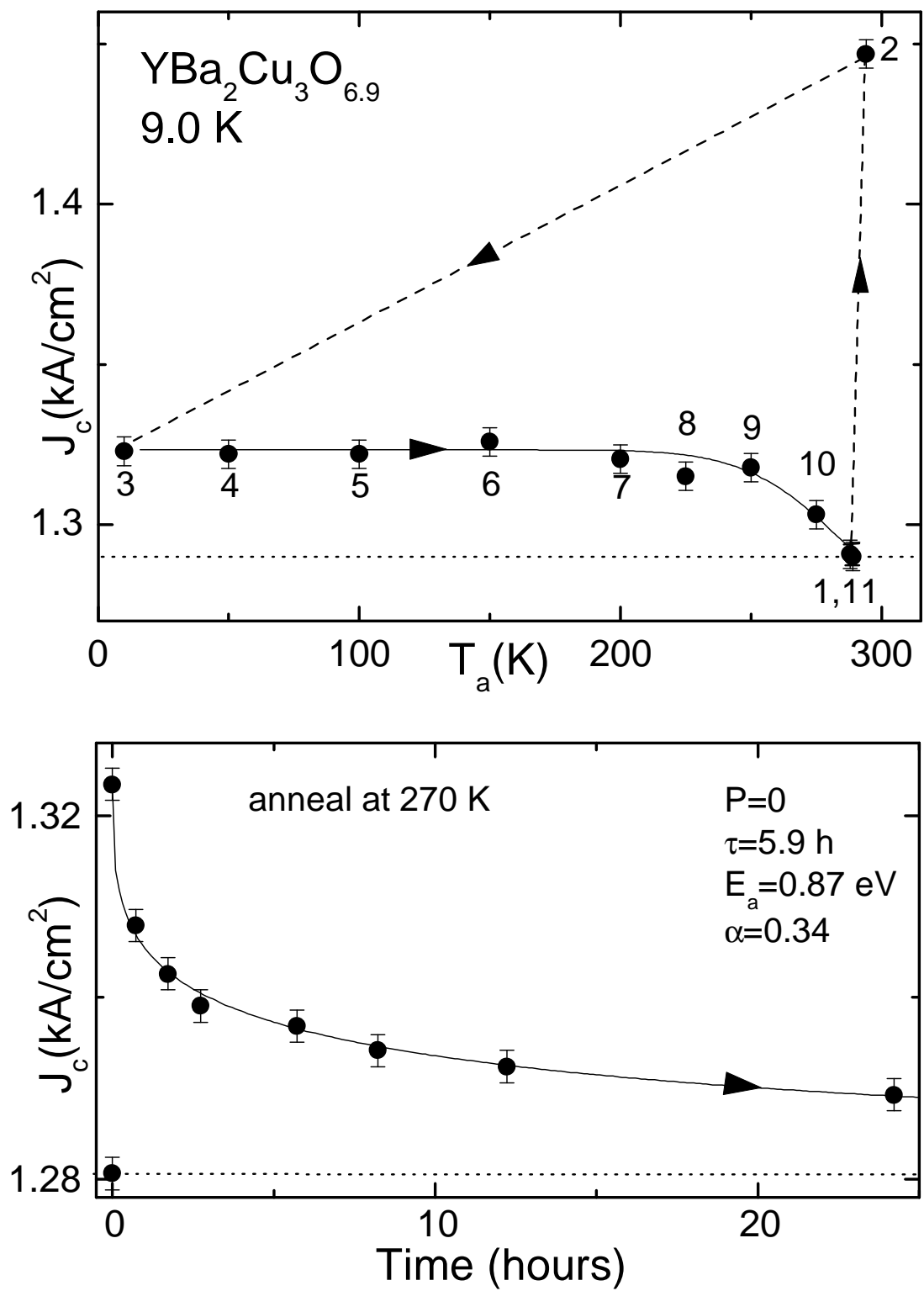
\noindent\textbf{Fig. 2.}\ \ For nearly optimally doped YBCO bicrystalline
ring with 25$^{\circ}$ GB: (top) dependence of critical current density
$J_{c}$ at 9 K on annealing temperature $T_{a}.$ Numbers give order of
measurement. Solid and dashed lines are guides to the eye. Horizontal dotted
line marks initial value of $J_{c}(9$ K) at point 1. (bottom) Dependence of
$J_{c}$ on time for fixed annealing temperature $T_{a}=$ 270 K. Solid line is
fit to data using Eq. (3).


\begin{thebibliography}{99}
\bibitem {note1}In view of the fact that GBs contain vacancies and other
structural defects, $\kappa_{GB}$\ is likely somewhat greater than $\kappa
_{a}$. In YBCO, for example, the average linear compressibility in the
CuO$_{2}$\ plane increases by 17\% if the oxygen concentration in the chains
is reduced by 35\% (see Ref \cite{jorgensen1}).
\end{thebibliography}
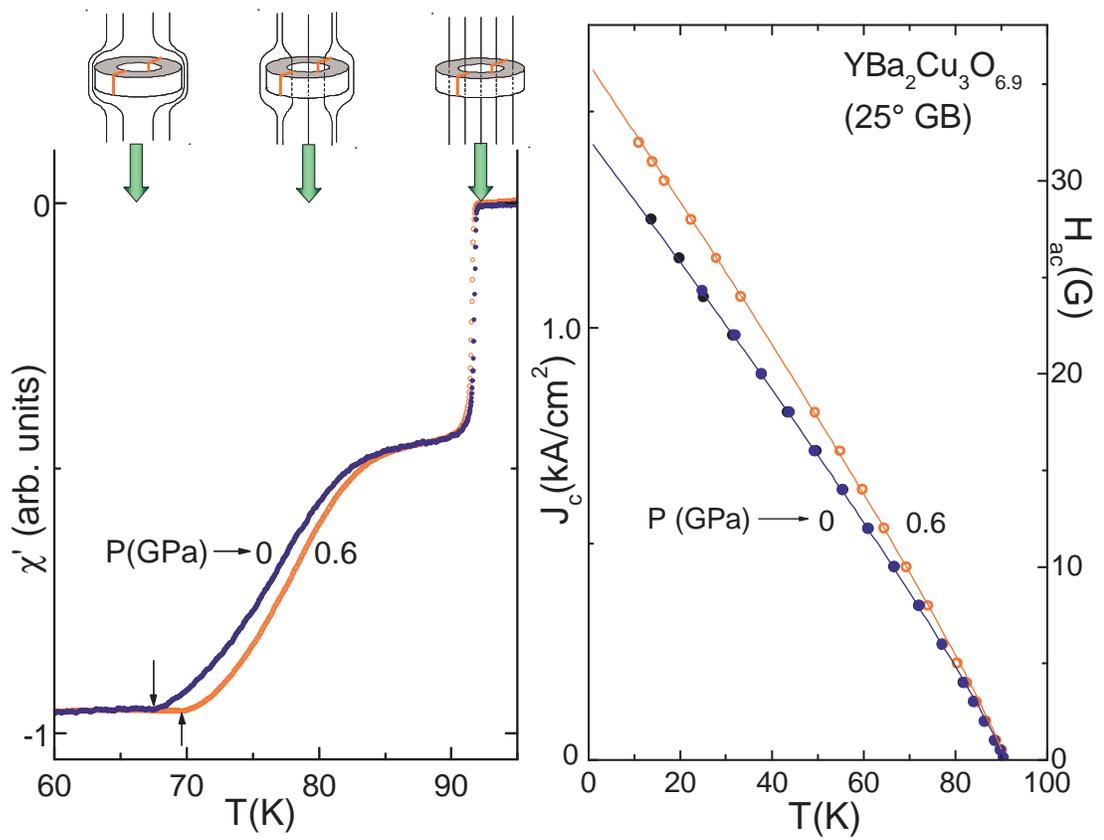
\end{document}